\documentclass[12pt]{iopart}
\usepackage{graphicx}
\usepackage{amssymb}
\usepackage{color}
\usepackage{setstack}
\usepackage{iopams}
\usepackage{soul}
\def\Msol{{M_\odot}}
\newcommand{\drond}[2][ ]{\frac{\partial #1}{\partial #2}}

\newcommand{\deriv}[2][ ]{\frac{\mathrm{d} #1}{\mathrm{d} #2}}

\renewcommand{\epsilon}{\varepsilon}

\def\CC{{C\nolinebreak[4]\hspace{-.05em}\raisebox{.4ex}{\tiny\bf ++} }}

\begin{document}

\title[New formulation of hydrodynamic equations]{A new formulation of general-relativistic hydrodynamic equations using primitive variables}

\author{Ga\"el Servignat$^{1}$, J\'er\^ome Novak$^{1}$ and Isabel Cordero-Carri\'on$^{2}$}
\address{$^{1}$ Laboratoire Univers et Th\'eories, Observatoire de Paris, Universit\'e PSL, CNRS, Universit\'e Paris-Cit\'e, 92190 Meudon, France\\
$^{2}$ Departamento de Matem\'aticas, Universitat de Val\`encia, 46100 Burjassot, Val\`encia, Spain} 

\begin{abstract}
   We present the derivation of hydrodynamical equations for a perfect fluid in General Relativity, within the 3+1 decomposition of spacetime framework, using only primitive variables. Primitive variables are opposed to conserved variables, as defined in the widely used Valencia formulation of the same hydrodynamical equations. The equations are derived in a covariant way, so that they can be used to describe any configuration of the perfect fluid. Once derived, the equations are tested numerically. We implement them in an evolution code for spherically symmetric self-gravitating compact objects. The code uses pseudospectral methods for both the metric and the hydrodynamics. First, convergence tests are performed, then the frequencies of radial modes of polytropes are recovered with and without the Cowling approximation, and finally the performance of our code in the black hole collapse and migration tests are described. The results of the tests and the comparison with a reference core-collapse and neutron star oscillations code suggests that not only our code can handle very strong gravitational fields, but also that this new formulation helps gaining a significant amount of computational time in hydrodynamical simulations of smooth flows in General Relativity.
\end{abstract}

\maketitle


\section{Introduction}
General-relativistic hydrodynamics is a key ingredient in many astrophysical models, most often combined with a numerical approach to obtain a solution (see~\cite{font_numerical_2008} and~\cite{rezzolla_relativistic_2013} for reviews on the subject). As examples, one can think of several phenomena in the field of high-energy astrophysics, such as binary neutron star (NS) mergers~\cite{baiotti_binary_2017}, NS oscillations~\cite{krueger_fast_2020}, core-collapse supernovae~\cite{janka_physics_2016} or accretion flows around black holes~\cite{abramowicz_foundations_2013}. These simulations have undergone tremendous progress in the last two decades, particularly since the first three-dimensional fully-relativistic code has simulated the merger of a binary NS system~\cite{shibata_simulation_2000}. Later major improvements have been obtained in the numerical modeling thanks to the development of so-called upwind high-resolution shock-capturing schemes written in conservation form and based on the characteristic fields of the system of hydrodynamics equations~\cite{banyuls_numerical_1997}. This type of methods possesses the great advantage of being able to sharply and stably resolve shocks and is thus able to model ultrarelativistic flows. However, the fact of using conservation schemes means that time evolution is done with \emph{conserved\/} variables, which are different from \emph{primitive\/} variables (velocity and thermodynamic quantities). Therefore, it is necessary, at each time-step, to pass from conserved to primitive variables with the solution of a set of equations and the call to the equation of state (EoS). Although some optimization is possible~\cite{noble_type_2008}, this step requires a non-negligible amount of computing time and may be the source of code failure, in particular if the EoS is not given by an analytic expression. Note that Lagrangian methods often rely on this recovery in their algorithm too~\cite{rosswog_sphincs_bssn_2021}. 

In contrast to the ultrarelativistic cases mentioned above, numerical studies of subsonic flows may not require such sophisticated approaches and, in order to speed-up the numerical process, an interesting alternative can be to evolve directly the primitive variables, not considering the hydrodynamic equations in a conserved form. By doing so, the recovery step is skipped, and the potential problems associated with it are avoided. The aim of this article is to derive such explicit hydrodynamics equations for primitive variables, within the so-called 3+1 formalism of General Relativity, and to test them in some simplified setting (one-dimensional fluid with analytic EoS) in the simulation of the evolution of a self-gravitating compact star. The paper is organized as follows: in Section~\ref{s:fluid_variables}, we introduce the fluid variable notations and thermodynamic properties, and the general four-dimensional covariant equations; Section~\ref{s:hydro_equations} is devoted to the computation of hydrodynamic equations in their 3+1 form, using explicitly primitive variables; these equations are then implemented in a numerical code, presented in Section~\ref{s:code_description} with which several tests are performed to compare the new formulation and code with previous ones, and are presented in Section~\ref{s:numerical_tests}. Finally, Section~\ref{s:conclusions} summarizes the results and gives some concluding remarks. In the following, we use units such that $c=1$, a 4-metric $g_{\mu\nu}$ with a signature $\left(-,+,+,+\right)$, and a covariant derivative (connection) $\nabla_\mu$ associated to that metric. Greek indices are running from 0 to 3, whereas Latin lowercase ones are running from 1 to 3. Latin uppercase indices are used to denote various thermodynamic species present in the fluid.

\section{Fluid variables and covariant conservation laws}\label{s:fluid_variables}
\subsection{Notations and definitions}\label{subs:notations}
We consider a perfect fluid, for which the energy-momentum tensor is:
\begin{equation}\label{eq:tmunu}
	T_\mathrm{fluid}^{\mu\nu} = (e+p)u^\mu u^\nu + pg^{\mu\nu}
,\end{equation}
where $u^\mu$ is the timelike unitary 4-velocity that carries all species and that satisfies $u_\mu u^\mu = -1$.
$e$ is the energy density in the fluid frame, and $p$ the pressure. The thermodynamics of the fluid is described with the following parameters: the number densities $n_X$ where $X$ spans all the types of species in the fluid, with the particular case of the subscript $B$ that refers to the baryons, and the entropy density $S$ which is conjugate to the temperature $\theta$: 
\begin{equation}
	\theta=\theta(S,n_X)=\left.\drond[e(S,n_X)]{S}\right|_{n_X}.
\end{equation}
We shall denote by $\hat{S}=S/n_B$ the entropy per baryon and $Y_A=n_A/n_B$ the abundancies, where $A$ is a subscript that refers to any species except the baryons. Finally let
\begin{equation}
    \mu_Z=\mu_Z(S,n_X)=\left.\drond[e(S,n_X)]{n_Z}\right|_{S,n_{Y\neq Z}}
\end{equation}
be the chemical potential associated with the species $Z$.

\subsection{Linking the thermodynamic variables}
The differential of the $e(S,n_X)$ is:
\begin{equation}\label{eq:diffe}
	\rmd e=\theta\,\rmd S + \sum\limits_Z \mu_Z\,\rmd n_Z
.\end{equation}
The relation for the pressure $p$ as a function of $S$ and $n_X$ reads:
\begin{equation}\label{eq:defp}
	p=p(S,n_X)=\theta S + \sum\limits_Z \mu_Z n_Z - e.
\end{equation}
The differential of $p(S,n_X)$ is:
\begin{equation}\label{eq:diffp}
    \rmd{p} = \left.\drond[p]{S}\right|_{n_X}\rmd{S} + \sum\limits_Z\left.\drond[p]{n_Z}\right|_{S,n_{Y\neq{Z}}}\rmd{n_Z}
.\end{equation}
The definition of the squared sound velocity is given e.g. by~\cite{typel_compose_2015} :
\begin{equation}\label{eq:csound}
    c_s^2 = \left.\deriv[p]{e}\right|_{\rmd{s}=\rmd{N_B}=\rmd{N_A}=0},
\end{equation}
where $s=S\,V$, $N_X=n_XV$ are the entropy and number of X particles, with $V$ the volume.
Eq.~(\ref{eq:csound}) can be rewritten as:
\begin{equation}\label{eq:cs2}
	c_s^2 = \frac{1}{e+p}\left(\left.\drond[p]{S}\right|_{n_X}S + \sum\limits_Z \left.\drond[p]{n_Z}\right|_{S,n_{Y\neq Z}} n_Z\right)
.\end{equation}

\subsection{Relativistic conservation laws}
The conservation of energy and momentum is written with the following Ansatz for the stress-energy tensor:
\begin{equation}
	\nabla_\mu(T_\mathrm{fluid}^{\mu\nu} + T_\mathrm{emission}^{\mu\nu}) = 0,
	\label{eq:tmunuconserv}
\end{equation}
where $T_\mathrm{fluid}^{\mu\nu}$ is defined by Eq.~(\ref{eq:tmunu}) and $T_\mathrm{emission}^{\mu\nu}$ encodes all possible emissions such as electromagnetic radiation or neutrino emission \textit{via} weak reactions. The latter can be incorporated as bulk viscosity~\cite{camelio_simulating_2022}. The equation is then rewritten as:
\begin{equation}
    \nabla_\mu T_\mathrm{fluid}^{\mu\nu} = Q^\nu
,\end{equation}
denoting $Q^\nu = -\nabla_\mu T_\mathrm{emission}^{\mu\nu}$. The conservation of a species $X$ flux is, with a possible source term $\sigma_X$ that account for possible processes such as chemical or nuclear reactions:
\begin{equation}
	\nabla_\mu(n_Xu^\mu)=\sigma_X
	\label{eq:baryonconserv}
.\end{equation}
Projecting Eq.~(\ref{eq:tmunuconserv}) along $u_\nu$ 
gives, also denoting $Q=u_\nu Q^\nu$:
\begin{equation}
	\theta\,\nabla_\mu(S\,u^\mu) = -Q - \sum\limits_Z\mu_Z\sigma_Z.
\end{equation}
Then, in the case $\theta\neq0$:
\begin{equation}\label{eq:Sconserv}
	\nabla_\mu(S\,u^\mu) = -\frac{1}{\theta}\left(Q+\sum\limits_Z\mu_Z\sigma_Z\right)
,\end{equation}
and in the case $\theta=0$ the entropy density in the fluid is also 0 as stated by the third law of thermodynamics, and the evolution equation for $S$ is irrelevant.

\subsection{Substitution  of pressure gradients with the log-enthalpy}\label{subs:enthalpysubstitution}
As we shall see in Secs.~\ref{subs:evoleqs} and~\ref{subs:evoleqsnum}, Eqs.~(\ref{eq:evolS}),~(\ref{eq:evolnx}),~(\ref{eq:evolUi}) and~(\ref{eq:evolH}) involve the ratio between the gradient of $p$ and $e+p$, which is numerically ill-defined for applications to self-gravitating objects, as both terms can vanish near the border of such an object. Let $H$ be the log-enthalpy:
\begin{equation}
    H = \ln\left(\frac{e+p}{m_Bn_B}\right), 
\end{equation}
with $m_B$ a baryon mass. Then:
\begin{equation}\label{eq:diffH}
	\frac{\rmd{p}}{e+p} = \rmd{H} -  e^{-H}\left(\theta\rmd\hat{S}+\sum_{A\neq B}\mu_A\rmd{Y_A}\right)
,\end{equation}
where $B$ denotes the baryons. We thus see that the problematic terms can be replaced with a well-behaved expression that does not involve any ratio of vanishing quantities.

\section{Hydrodynamic equations in 3+1 form}\label{s:hydro_equations}
\subsection{3+1 decomposition}
In this subsection, we introduce the notations that we use within the 3+1 formalism framework. For a full introduction to 3+1 formalism and 3+1 hydrodynamics, the reader is referred to~\cite{gourgoulhon_31_2012} and references therein.

We consider an asymptotically flat spacetime. The spacetime metric is given in the standard 3+1 form:
\begin{equation}
	g_{\mu\nu}\mathrm{d}x^\mu\mathrm{d}x^\nu := -N^2\mathrm{d}t^2 + \gamma_{ij}(\mathrm{d}x^i + \beta^i\mathrm{d}t)(\mathrm{d}x^j + \beta^j\mathrm{d}t)
,\end{equation}
where $N$ is the lapse function, $\beta^i$ the shift 3-vector and $\gamma_{ij}$ the induced 3-metric on spacelike 3-hypersurfaces, associated with its covariant derivative $D_i$. The {extrinsic curvature} tensor is defined by
\begin{equation}
    K_{ij} := -\frac{1}{2}\mathcal{L}_\mathbf{n}\gamma_{ij}
,\end{equation}
where $\mathcal{L}_\mathbf{n}$ is the Lie derivative along the vector $n^\mu$, a unitary, future-oriented, timelike 4-vector normal to a hypersurface $\Sigma_t$ of constant time $t$:
\begin{equation}
    n_\mu := N\nabla_\mu{t}.
\end{equation}
$K:=K^i_i$ denotes the trace of the extrinsic curvature tensor. Finally, we consider the following hydrodynamical quantities: the Lorentz factor with respect to the Eulerian observer will be denoted $\Gamma$, the Eulerian velocity $U^\mu$ and the coordinate velocity $v^i$. These quantities are related according to:
\begin{equation}
	\Gamma := -n_\mu u^\mu = (1-U_iU^i)^{-1/2}
,\end{equation}
and
\begin{equation}
	u^\alpha = \Gamma(n^\alpha + U^\alpha) = \frac{\Gamma}{N}(1,v^i),\quad U^i=\frac{1}{N}(v^i+\beta^i)
.\end{equation}

\subsection{3+1 formulation of conservation laws}
In this subsection we write the conservation laws~(\ref{eq:baryonconserv}) and~(\ref{eq:Sconserv}) in the framework of the 3+1 formalism, and write the 3+1 relativistic equivalent of Euler's equation. Rewriting Eq.~(\ref{eq:Sconserv}) by reformulating the divergence with $\gamma=\det(\gamma_{ij})$, the determinant of the 3-metric:
\begin{equation}
	\nabla_\mu(S\,u^\mu) = \frac{1}{N\sqrt{\gamma}}\partial_\mu\left(N\sqrt{\gamma}\,S\,u^\mu\right) = -\frac{1}{\theta}\left(Q+\sum\limits_Z\mu_Z\sigma_Z\right)
.\end{equation}
Using $\Gamma=Nu^0$ and then the 3-dimensional divergence formula, the previous equation is written as:
\begin{equation}
	\partial_t(S\,\Gamma\sqrt{\gamma}) + \sqrt{\gamma}D_i(S\,\Gamma\,v^i) = -\frac{N\sqrt{\gamma}}{\theta}\left(Q+\sum\limits_Z\mu_Z\sigma_Z\right)
.\end{equation}
Expanding the time derivative we get:
\begin{equation}\label{eq:conservationentropy3plus1}
	\partial_t S = -S\frac{\partial_t\Gamma}{\Gamma} - S \frac{\partial_t\gamma}{2\gamma} - \frac{D_i(S\,\Gamma\, v^i)}{\Gamma} - \frac{N}{\Gamma\theta}\left(Q+\sum\limits_Z\mu_Z\sigma_Z\right)
.\end{equation}
Then we can deduce analogously that:
\begin{equation}\label{eq:conservationbaryon3plus1}
	\partial_t n_X = -n_X\frac{\partial_t\Gamma}{\Gamma} - n_X \frac{\partial_t\gamma}{2\gamma} - \frac{D_i(n_X\Gamma v^i)}{\Gamma} + \frac{N}{\Gamma}\sigma_X
.\end{equation}
The 3+1 Euler's equation is~\cite{gourgoulhon_31_2012}:
\begin{eqnarray}\label{eq:euler3plus1}
	\partial_t U_i +v^jD_jU_i & = - \frac{1}{\Gamma^2(e+p)}\left[ND_i p + U_i(\partial_tp - \beta^jD_jp)\right] \nonumber \\
							  & + U_jD_i\beta^j - D_iN + U_iU^j(D_jN - NK_{jk}U^k)
.\end{eqnarray}
As we can see, the evolution of the parameters is governed by the evolution of the Lorentz factor, the determinant of the 3-metric and the pressure. The following subsections will be dedicated to the derivation of the evolution equations for those three quantities. At the end, the equations will be decoupled to have evolution equations for $S,\,n_X$ and $U_i$ that only rely on spatial derivatives of the thermodynamical and hydrodynamical primitive quantities.

\subsection{Evolution equation for $\gamma$}
The evolution equation for the 3-metric combined with the derivative of a determinant yields~\cite{gourgoulhon_31_2012}:
\begin{equation}\label{eq:evolgamma}
	\frac{\partial_t\gamma}{2\gamma} = - NK + D_i\beta^i
.\end{equation}

\subsection{Evolution equation for $p$}\label{sec:evolp}
Starting from Eq.~(\ref{eq:diffp}) and using Eqs.~(\ref{eq:conservationentropy3plus1}),~(\ref{eq:conservationbaryon3plus1}) and~(\ref{eq:cs2}):
\begin{eqnarray}
	\partial_tp & = \left.\drond[p]{S}\right|_{n_X}\partial_tS + \sum\limits_Z\left.\drond[p]{n_Z}\right|_{S,n_{Y\neq Z}}\partial_tn_Z\nonumber \\
	& = -(e+p)c_s^2\left(\frac{\partial_t\Gamma}{\Gamma} + \frac{\partial_t\gamma}{2\gamma} + \frac{D_i(\Gamma v^i)}{\Gamma}\right) - v^iD_ip \nonumber \\
				& + \frac{N}{\Gamma}\left(-\left.\drond[p]{S}\right|_{n_X}\frac{1}{\theta}\left(Q+\sum\limits_Z\mu_Z\sigma_Z\right) + \sum\limits_Z\left.\drond[p]{n_Z}\right|_{S,n_{Y\neq Z}}\sigma_Z\right)
				\label{eq:evolp}
.\end{eqnarray}
The last term in the equation is computed from the EoS and the microphysical reaction rates. In the following, it will be denoted with $\Xi$:
\begin{equation}
    \Xi = \left(-\left.\drond[p]{S}\right|_{n_X}\frac{1}{\theta}\left(Q+\sum\limits_Z\mu_Z\sigma_Z\right) + \sum\limits_Z\left.\drond[p]{n_Z}\right|_{S,n_{Y\neq Z}}\sigma_Z\right)
.\end{equation}
Note that in the special case of no emission ($Q=0$) and no chemical or nuclear reactions ($\sigma_X=0, \forall X$), $\Xi=0$.

\subsection{Evolution equation for $\Gamma$}
The evolution equation for the {Lorentz} factor is deduced from its relation to $U_i$:
\begin{equation}
	\partial_t\Gamma = \frac{1}{2}\Gamma^3\partial_t(\gamma^{ij}U_iU_j) = \frac{1}{2}\Gamma^3(U_iU_j\partial_t\gamma^{ij} +  2U^i\partial_tU_i)
. \end{equation}
Injecting Eqs.~(\ref{eq:euler3plus1}) and~(\ref{eq:evolp}), as well as the evolution equation for the inverse metric~\cite{gourgoulhon_31_2012} in the previous equation, gives:
\begin{eqnarray}
	\partial_t\Gamma & = \Gamma(NU^kU^lK_{kl} - U^jD_jN) -\Gamma^3U^iv^jD_jU_i \nonumber \\
					 & - \frac{\Gamma U^i}{e+p}\bigg[ND_ip + U_i\bigg(- v^jD_jp + \frac{N}{\Gamma}\Xi - \beta^jD_jp  \nonumber \\
					 & \left.\left.-(e+p)c_s^2\left(\frac{\partial_t\Gamma}{\Gamma} + \frac{\partial_t\gamma}{2\gamma} + \frac{D_i(\Gamma v^i)}{\Gamma}\right)\right)\right] 
					 \label{eq:evolGamma}
.\end{eqnarray}
Bringing all the $\partial_t\Gamma$ terms together on the left hand side finally yields:
\begin{eqnarray}
	\frac{\partial_t\Gamma}{\Gamma} & = \frac{\Gamma^2}{\Gamma^2-c_s^2(\Gamma^2-1)}\Biggl[NU^iU^jK_{ij} - U^iD_iN - \Gamma^2U^iv^jD_jU_i - \frac{\Gamma^2-1}{\Gamma^2(e+p)}\frac{N}{\Gamma}\Xi \nonumber \\
	& - \frac{1}{\Gamma^2(e+p)}NU^jD_jp  + c_s^2\left(\frac{\Gamma^2-1}{\Gamma^2}\right)\left(\frac{\partial_t\gamma}{2\gamma} + \frac{D_i(\Gamma v^i)}{\Gamma}\right)\Biggl]
.\end{eqnarray}

\subsection{Evolution equations for $S,\,n_X$ and $U_i$}\label{subs:evoleqs}
Replacing Eqs.~(\ref{eq:evolgamma}) and~(\ref{eq:evolGamma}) in Eqs.~(\ref{eq:conservationentropy3plus1}) and~(\ref{eq:conservationbaryon3plus1}) gives the evolution equations for $S$ and $n_X$:
\begin{eqnarray}\label{eq:evolS}
	\partial_tS & = -v^iD_iS - \frac{SN\Gamma^2}{\Gamma^2-c_s^2(\Gamma^2-1)}\Bigg[U^iU^jK_{ij} - K - \frac{U^jD_jp}{\Gamma^2(e+p)} + D_jU^j \nonumber\\
	&  - \frac{\Gamma^2-1}{\Gamma^2(e+p)}\frac{\Xi}{\Gamma}\Bigg]  - \frac{N}{\Gamma}\frac{1}{\theta}\left(Q+\sum\limits_Z\mu_Z\sigma_Z\right), \\
	\label{eq:evolnx}
	\partial_tn_X & = -v^iD_in_X - \frac{n_XN\Gamma^2}{\Gamma^2-c_s^2(\Gamma^2-1)}\Bigg[U^iU^jK_{ij} - K - \frac{U^jD_jp}{\Gamma^2(e+p)} + D_jU^j \nonumber\\
	& - \frac{\Gamma^2-1}{\Gamma^2(e+p)}\frac{\Xi}{\Gamma}\Bigg] + \frac{N}{\Gamma}\sigma_X
.\end{eqnarray}
For the evolution equation for $U_i$, we use Eqs.~(\ref{eq:evolp}), (\ref{eq:evolgamma}) and~(\ref{eq:evolGamma}): 

\begin{eqnarray}\label{eq:evolUi}
\partial_tU_i
    & = -v^jD_jU_i  + U_jD_i\beta^j - D_iN + U_iU^jD_jN \nonumber \\
	& + \frac{c_s^2\, N U_i}{\Gamma^2-c_s^2(\Gamma^2-1)}(D_jU^j-K) + U_i\frac{\Gamma^2(c_s^2-1)}{\Gamma^2-c_s^2(\Gamma^2-1)}NU^lU^jK_{lj} \nonumber\\
	&
	- \frac{N}{\Gamma^2}\left(\frac{D_ip}{e+p} - \frac{\Gamma^2(1-c_s^2)}{\Gamma^2-c_s^2(\Gamma^2-1)}U_iU^j\frac{D_jp}{e+p}\right) \nonumber\\
	& - \frac{U_i}{\Gamma^2-c_s^2(\Gamma^2-1)}\frac{N\Xi}{\Gamma(e+p)}
.\end{eqnarray}
Note that the causality condition is $c_s^2\leq1$, and $\Gamma^2\geq 1$ by definition, thus $\Gamma^2-c_s^2(\Gamma^2-1)\geq 1$, and therefore the denominator $\Gamma^2-c_s^2(\Gamma^2-1)$ never vanishes.

These equations can be analytically compared with some particular cases: we checked that in the Newtonian limit, in a non reactive system (i.e. with $Q=0$ and $\sigma_X=0, \forall X$), Eqs.~(\ref{eq:evolnx}) and~(\ref{eq:evolUi}) yield their Newtonian counterparts, and that using the polar slicing and radial gauge, these equations exactly correspond to that of~\cite{gourgoulhon_simple_1991}.

\subsection{Set of evolution equations for numerical simulations}\label{subs:evoleqsnum}
Depending on the different equilibrium conditions, the EoS depends on a certain number of thermodynamical variables, that we call parameters. For practical applications, the number of parameters in the EoS is usually between one and three. The choice of three parameters typically corresponds to the description of a hot fluid where weak nuclear $\beta$ reactions take place. The three parameters are usually $(\theta,n_B,Y_e)$, or any thermodynamically equivalent quantity. $Y_e$ corresponds to the electron fraction in the fluid. However, the density of baryons may be discontinuous across a perfect fluid, for example in the presence of a phase transition, whereas the log-enthalpy is better-behaved. For applications using pseudospectral methods, see Sec.~\ref{subs:numericalimplementation}, we use $(\hat{S},H,Y_e)$, where those parameters were defined in Secs.~\ref{subs:notations} and~\ref{subs:enthalpysubstitution}. The source term for the baryon density is $\sigma_B := 0$. It encodes that the baryon number is strictly conserved. The conservation of the lepton number is:
\begin{equation}
    \nabla_\mu(n_lu^\mu) = 0
.\end{equation}
Therefore, the associated chemical potential is that of the leptons:
\begin{equation}
    \mu_l := \left.\drond[e]{n_l}\right|_{S,n_B}.
\end{equation}
However, we assume that the fluid is transparent to neutrinos so that they escape instantly. Under this particular assumption, with $n_l = n_e + n_{\nu_e}$, the energy density does not contain the contribution from the neutrinos, and the lepton chemical potential expression is rewritten as:
\begin{equation}
    \mu_l = \left.\drond[e]{n_e}\right|_{S,n_B}
,\end{equation}
and denoting $\sigma = \sigma_e := -\nabla_\mu(n_{\nu_e}u^\mu)$ the conservation law becomes:
\begin{equation}
    \nabla_\mu(n_eu^\mu) = \sigma
.\end{equation}
Then, the knowledge of $\partial_tS$, $\partial_tn_B$ and $\partial_tn_e$ with Eqs.~(\ref{eq:evolS}),~(\ref{eq:evolnx}), allows to compute $\partial_t\hat{S},\,\partial_tH$ and $\partial_tY_e$ from the differential relation between those variables. Let us start with the electron fraction:
\begin{equation}
	\rmd{Y_e} = \frac{1}{n_B}\rmd{n_e} - \frac{Y_e}{n_B}\rmd{n_B}
.\end{equation}
Then the evolution equation for $Y_e$ is simply:
\begin{equation}\label{eq:evolYe}
	\partial_t Y_e = -v^iD_iY_e + \frac{N}{\Gamma}\frac{\sigma}{n_B}
.\end{equation}
Then, for the entropy per baryon, we relate the differentials of $S$ and $\hat{S}$:
\begin{equation}
	\rmd{\hat{S}} = \frac{1}{n_B}\left(\rmd{S} - \hat{S}\rmd{n_B}\right)
.\end{equation}
The evolution equation for $\hat{S}$ is then once again a simple advection equation:
\begin{equation}\label{eq:evolhatS}
	\partial_t\hat{S} = -v^iD_i\hat{S} - \frac{N}{\Gamma\theta n_B}\left(Q + \mu_l\sigma\right)
.\end{equation}
Those two equations are the same as the ones of~\cite{pascal_proto-neutron_2022}, that were also derived by~\cite{pons_evolution_1999}. The corresponding differential of $H$ in the particular case of the parameters ($\hat{S},H,Y_e$) that we chose is:
\begin{equation}
	\rmd{H} = \frac{\rmd{p}}{e+p} + e^{-H}(\theta\rmd{\hat{S}} + \mu_l\rmd{Y_e})
,\end{equation}
Therefore, combining Eqs.~(\ref{eq:evolp}),~(\ref{eq:evolYe}) and~(\ref{eq:evolhatS}), the evolution equation for the log-enthalpy is:
\begin{eqnarray}\label{eq:evolH}
	\partial_tH & = -v^iD_iH - c_s^2\frac{\Gamma^2N}{\Gamma^2 - c_s^2(\Gamma^2-1)}\bigg[K_{j}U^iU^j - K + D_iU^i \nonumber \\
				&  - \frac{U^i}{\Gamma^2}\left(D_iH -e^{-H}\left(\theta D_i\hat{S} + \mu_lD_iY_e\right)\right)\bigg] + e^{-H}\frac{N}{\Gamma}\frac{Q}{n_B} \nonumber\\
			    & + e^{-H}\frac{N}{\Gamma}\frac{(\Gamma^2-1)c_s^2}{\Gamma^2 - c_s^2(\Gamma^2-1)}\Bigg(\left.\drond[p]{n_e}\right|_{S,n_B}\frac{\sigma}{n_B} - \left.\drond[p]{S}\right|_{n_e,n_B}\frac{Q+\mu_l\sigma}{n_B\theta}\Bigg)
.\end{eqnarray}
The final set of variables that are evolved through partial differential equations are $(U_i,H,Y_e,\hat{S})$, for which the corresponding equations are Eqs.~(\ref{eq:evolUi}),~(\ref{eq:evolH}),~(\ref{eq:evolYe}) and~(\ref{eq:evolhatS}), where all the terms of the form $D_ip/(e+p)$ have been replaced thanks to Eq.~(\ref{eq:diffH}). The system must be closed by providing an EoS, which links the pressure, energy density, sound speed, chemical potential and temperature to the parameters $(H,Y_e,\hat{S})$. In order to check the validity of the set of equations, we will perform numerical tests that will be described in the following sections.

\section{Description of the code}\label{s:code_description}
\subsection{Conformal decomposition and choice of coordinates}
In order to fix the coordinates, we must choose a gauge and a foliation. Following~\cite{bonazzola_constrained_2004}, we use the conformal decomposition of spacelike hypersurfaces. Let $f_{ij}$ be a flat metric, $\bar{D}_i$ its associated covariant derivative, $f = \det(f_{ij})$. We require that $\gamma_{ij}$ is $f_{ij}$ at spatial infinity. By introducing the so-called conformal factor
\begin{equation}
	\Psi = \left(\frac{\gamma}{f}\right)^{1/12}
,\end{equation}
the conformal metric is defined as
\begin{equation}
	\tilde{\gamma}_{ij} = \Psi^{-4}\gamma_{ij}
.\end{equation}
We choose the maximal slicing as a foliation, mathematically expressed as:
\begin{equation}
	K = 0
.\end{equation}
This choice has the effect of simplifying some of the equations and it numerically avoids possible black hole singularities, for example during a gravitational collapse, see Sec.~\ref{subs:blackholecollapse}. We also choose the {Dirac} gauge:
\begin{equation}
	\bar{D}_i\tilde{\gamma}^{ij} = 0
.\end{equation}
In spherical symmetry, it corresponds to the isotropic gauge and $\tilde{\gamma}^{ij} = f^{ij}$. The {Dirac} gauge is straightforward to use in three dimensions and allows to compute quasi-stationary initial data~\cite{bonazzola_constrained_2004}. 
\subsection{Numerical implementation in spherical symmetry}\label{subs:numericalimplementation}
We will perform various numerical tests on a one-species, spherically symmetric, cold, catalysed (i.e. where $\beta$ reactions are at equilibrium) NS, which is described by a one-parameter EoS. We choose the parameter to be the log-enthalpy $H$. In this case, the EoS only enters through the sound speed squared $c_s^2(H)$. In particular, $\Xi=Q=\sigma=0$ in Eqs.~(\ref{eq:evolUi}) and~(\ref{eq:evolH}). The evolution equations are that of $U_r$ (the first component of $U_i$ in a spherical triad)~(\ref{eq:evolUi}) and $H$~(\ref{eq:evolH}). We implement them along with the spherically symmetric version of the Einstein system of~\cite{cordero-carrion_improved_2009} (Eqs.~(30) to~(33) therein), which is a fully constrained formulation of Einstein equations, formulated as a set of four coupled Poisson-like partial differential equations.
The algorithm to perform the time evolution is the following:
\begin{itemize}
    \item At a given time $t$, $H$ and $U_r$ are known thanks to their evolution equations~(\ref{eq:evolH}) and~(\ref{eq:evolUi}). $\Gamma$ can be computed from $U_r$, and along with the EoS, $H$ allows to compute $p$ and $e$. $\Psi$ is temporarily known thanks to the following evolution equation, which can be obtained from Eq.~(\ref{eq:evolgamma}):
    \begin{equation}\label{eq:evolpsi}
        \partial_t\ln\Psi = \beta^i\bar{D}_i\ln\Psi + \frac{1}{6}\bar{D}_i\beta^i
    .\end{equation}
    It is then possible to compute the following quantities,
    \begin{eqnarray}
	    E^*   & = \Psi^6(\Gamma^2(e+p) - p),   \label{eq:estar}\\
	    S^* & = \Psi^6(3p + (\Gamma^2-1)(e+p)), \label{eq:Sstar}\\
	    p_r^* & = \Psi^6\Gamma^2(e+p)U_r, \label{eq:pjstar}
    \end{eqnarray}
    and solve the Einstein equations. The output is the value of $\Psi$ consistent with the Hamiltonian constraint, which replaces the value of $\Psi$ computed with Eq.~(\ref{eq:evolpsi}) as soon as it is available, as well as the lapse $N$ and the radial component of the shift vector $\beta^r$.
    \item We use the knowledge of the radius of the star $R$, as well as $H,\,U_r,\Psi,\,N$ and $\beta^r$ together with the EoS to compute the sources of Eqs.~(\ref{eq:evolUi}),~(\ref{eq:evolH}),~(\ref{eq:evolR}) and~(\ref{eq:evolpsi}). An explicit finite-difference time scheme allows to compute $\Psi,\,R,\,H$ and $U_r$ at the next time step.
\end{itemize}

The implementation is based upon LORENE~\cite{gourgoulhon_lorene_2016}, a dedicated \CC library that comes with built-in Poisson solvers, based on pseudospectral methods~\cite{grandclement_multidomain_2001,gottlieb_numerical_1977,boyd_chebyshev_2001,grandclement_spectral_2009}. We use the relaxation iterative method to solve the non-linear elliptic partial differential equations. Pseudospectral methods are also used to compute the spatial derivatives in the source of the evolution equations. The time evolution is done thanks to an explicit finite differences Adams-Bashforth scheme of order 3. The numerical grid comprises two domains: a nucleus centered on the star, with the domain boundary corresponding to the surface of the star, and a compactified external domain (CED) from the surface of the star to infinity, allowing to solve Poisson-like equations and imposing boundary conditions at spatial infinity. As the star is evolved, the boundary between the nucleus and the CED will move with it. This semi-Lagrangian approach was used in~\cite{gourgoulhon_simple_1991}. In order to do so, we perform a change of variables in the nucleus defined as follows:
\begin{equation}
	\left\{\begin{array}{ll}
		t'  & = t, \\
		\xi & = r/R(t).
	\end{array}\right.
\end{equation}
The radial coordinate is mapped from $[0,R(t)]$ to the numerical coordinate $\xi\in[0,1]$. This changes the differential operators with respect to $r$ and $t$:
\begin{eqnarray}
    \drond[]{r} & = \frac{1}{R}\drond[]{\xi}, \\
	\drond[]{t} & = -\frac{\dot{R}}{R}\xi\drond[]{\xi} + \drond[]{t'} \label{eq:chgtvarrxinuc}
,\end{eqnarray}
where $\dot{R}$ is the time derivative of the radius. The term $-\dot{R}\,\xi$ can be interpreted as a \textit{grid advection} velocity inside the star. The radius is an additional parameter, therefore we have to introduce a new equation:
\begin{equation}\label{eq:evolR}
    \dot{R}(t') := \deriv[R]{t'} = v^r(R(t'),t')
.\end{equation}
It corresponds to the impermeable boundary condition, which requires that no matter crosses the external border of the domain. In the CED, a similar change of variables is performed: 
\begin{equation}
	\left\{\begin{array}{ll}
		t'  & = t, \\
		\xi   & = 1 - 2R(t)/r.
	\end{array}\right.
\end{equation}
The radial coordinate is mapped from $[R(t),+\infty]$ to the numerical coordinate $\xi\in[-1,1]$. The differential operators are then also changed accordingly:
\begin{eqnarray}
    \drond[]{r} & = \frac{(\xi-1)^2}{2R} \drond[]{\xi}, \\
    \drond[]{t} & = \frac{\dot{R}}{R}(\xi-1)\drond[]{\xi} + \drond[]{t'}
    \label{eq:chgtvarrxiCED}
.\end{eqnarray} 
In what follows, we make no distinction between $t$ and $t'$ as they are identical. 

The boundary conditions between the domains for the Poisson-like equations are imposed by the Poisson solvers of LORENE: the continuity of the metric and its radial derivative is required. At infinity, we impose that $\Psi\rightarrow1,\,N\rightarrow1,\,\beta^r\rightarrow0$, as the metric should be asymptotically flat. For the evolution equations, the boundary conditions are the following: 
\begin{itemize}
    \item At the center of the star, i.e. $\xi=0$ in the nucleus, we impose the regularity of the fields.
    \item Eqs.~(\ref{eq:evolH}) and~(\ref{eq:evolUi}), within the assumption of subsonic flow, need one boundary condition at the border of the star, i.e. $\xi=1$ in the nucleus. The impermeable boundary condition is naturally imposed by choosing a comoving grid. It also helps with the stability of the code to enforce that $H$ is constant at the surface of the star.
    \item The communication between the nucleus and the CED for Eq.~(\ref{eq:evolpsi}) is done in an upwind fashion. We refer to~\ref{app:characteristic} for the details of the characteristic speed computation. No boundary condition needs to be imposed at $\xi=1$ in the CED as the characteristic speed there is $\beta^r(r=+\infty,t)=0$.
\end{itemize}
All initial data are computed with LORENE, using the solver described in~\cite{lin_rotating_2006} in spherical symmetry. 
\section{Numerical tests}\label{s:numerical_tests}
\subsection{Convergence}\label{subs:convergencetest}
Here we describe the convergence tests that we performed. We use a polytropic EoS with parameters $\gamma=2$,
$\kappa=100$ (in geometrized units where $G=c=1$, supplemented with $\Msol=1$) that was used by~\cite{font_three-dimensional_2002} to perform tests, with a central log-enthalpy of $H_c = 0.2279$. The convergence is measured on the conservation of the ADM and baryon masses. First, we fix the timestep $\Delta{t} = 3.34\times10^{-4}$ ms and vary $N_r$ the number of radial grid points between 5 and 65. We let the star at equilibrium (meaning that it oscillates only because of numerical noise) for a 1ms simulation. We find exponential convergence on the maximum of the relative error on the ADM mass with respect to the equilibrium value, as expected with pseudospectral methods. The error reaches a plateau around $10^{-10}$ at 25 grid points. 
To check the convergence of the time scheme, we use the same star, but this time the simulations were run up to 10 ms, we set $N_r=17$, the Cowling approximation was used, and a Gaussian enthalpy profile was added to the equilibrium, initial enthalpy profile as a perturbation. The timestep was varied between $10^{-4}$ and $3.34\times10^{-3}$ ms. We find an order 3 power-law convergence on the relative error of the baryon mass with respect to the initial equilibrium baryon mass, which is expected as the error for a finite difference scheme of order $n$ must behave as $\mathcal{O}(\Delta{t}^n)$.

\begin{table*}[t]
\caption{Comparing polytropic frequencies (fundamental mode, first and second overtone) of the code with the ones found in the literature. The frequencies were extracted from a 1 s simulation. For the first polytrope~\cite{font_three-dimensional_2002}, $\kappa$ is expressed in units of $G^3\Msol^2/c^4$, and for the second~\cite{hartle_slowly_1975}, $\kappa$ is expressed in $\mathrm{km}^{4/3}$, i.e. in geometrized units where $G=c=1$. The central energy density of the polytrope from~\cite{hartle_slowly_1975} is $e_c = 3.16\times10^{14}\mathrm{g}\cdot\mathrm{cm}^{-3}$. The corresponding spectrum for the star of~\cite{font_three-dimensional_2002} is given on the left panel of Fig.~\ref{fig:spectrumpoly}.}
\label{tab:freqpoly}
\resizebox{15.7cm}{!}{\begin{tabular}{lllllllll}
	\br
	        & $\kappa$ & $\gamma$ & $H_c \,[c^2]$  & $M \,[\Msol]$ & $R$ [km] & Fund. [kHz] & 1st ov. [kHz] & 2nd ov. [kHz]\\
	\mr
	{Font} et al.~\cite{font_three-dimensional_2002} & $100$ & $2$ & $0.2279$ & $1.4$ & $14.15$ & $1.450$ & $3.958$ & $5.935$ \\
	This work & $100$ & $2$ & $0.2279$ & $1.401$ & $14.16$ & $1.442$ & $3.954$ & $5.915$ \\
	Relative difference            & & & & & & $0.6 \%$ & $0.1 \%$ & $0.3 \%$ \\
	\mr
	{Hartle} \& {Friedman}~\cite{hartle_slowly_1975} & $7.308$ & $5/3$ & $6.720\times10^{-2}$ & $\times$ & $\times$ & $0.824$ & $1.94$ & $2.86$ \\
	This work                           & $7.308$ & $5/3$ & $6.720\times10^{-2}$ & $0.4866$ & $16.49$ & $0.823$ & $1.95$ & $2.86$ \\
	Relative difference            & & & & & & $0.1 \%$ & $0.5 \%$ & $0.0 \%$ \\
	\br
\end{tabular}}
\end{table*}
\begin{table*}[t]
\caption{Comparing polytropic frequencies (fundamental mode, first and second overtone) of the code with the ones found in the literature using the {Cowling} approximation. The frequencies were extracted from a 1 s simulation. $\kappa$ is expressed in units of $G^3\Msol^2/c^4$. The corresponding spectrum is given on the right panel of Fig.~\ref{fig:spectrumpoly}.}
\label{tab:freqpolycowling}
\resizebox{15.7cm}{!}
{\begin{tabular}{lllllllll}
	\br
	        & $\kappa$ & $\gamma$ & $H_c \,[c^2]$  & $M \,[\Msol]$ & $R$ [km] & Fund. [kHz] & 1st ov. [kHz] & 2nd ov. [kHz] \\
	\mr
	{Font} et al.~\cite{font_three-dimensional_2002} & $100$ & $2$ & $0.2279$ & $1.4$ & $14.15$ & $2.696$ & $4.534$ & $6.346$ \\
	This work & $100$ & $2$ & $0.2279$ & $1.401$ & $14.16$ & $2.685$ & $4.548$ & $6.339$ \\
	Relative difference            & & & & & & $0.4 \%$ & $0.3 \%$ & $0.1 \%$ \\
	\br 
\end{tabular}}
\end{table*}

\begin{figure}
	\centering
	\begin{minipage}{0.51\textwidth}
	\centering
	\includegraphics[width=\columnwidth]{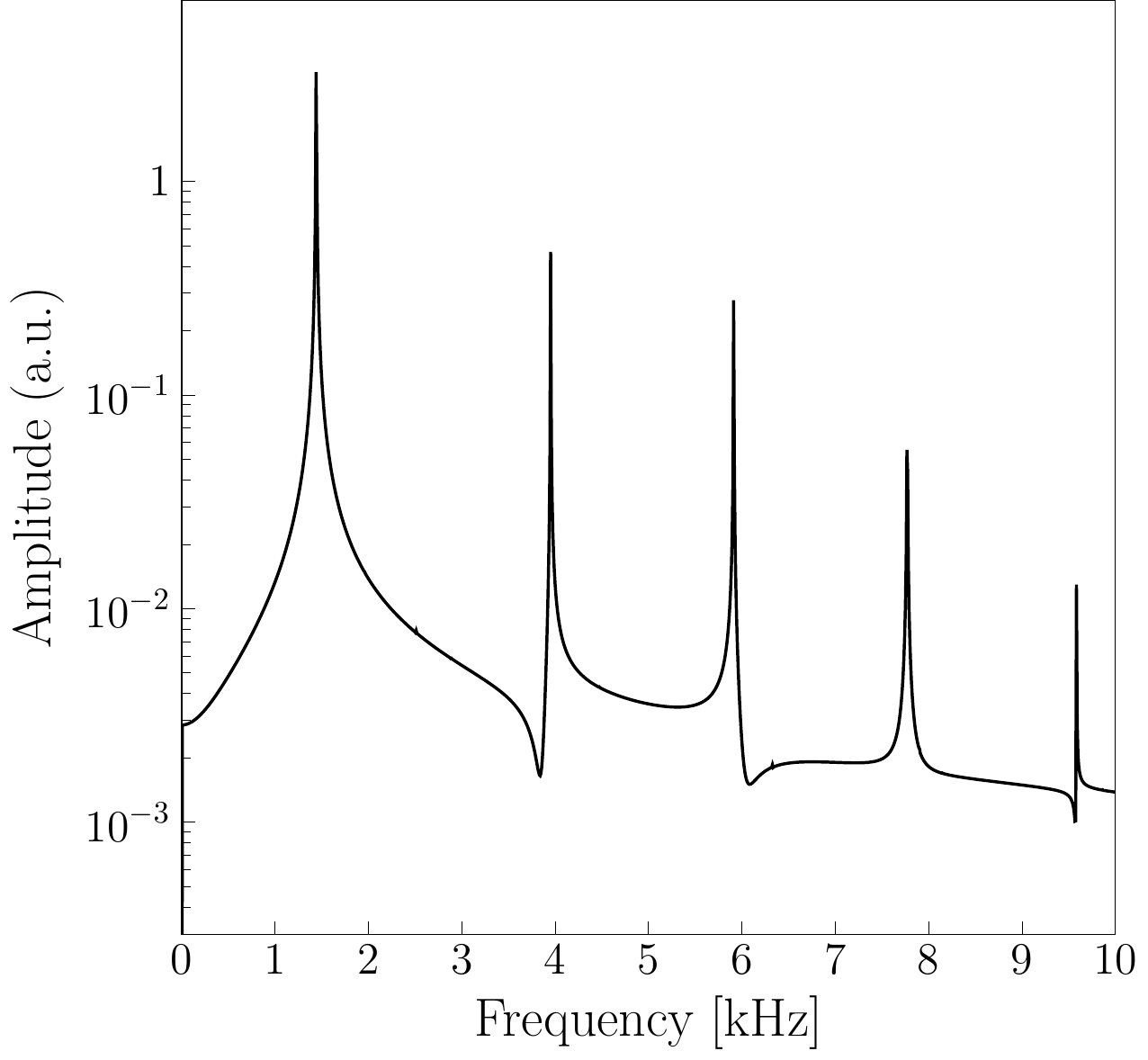}
	\end{minipage}
	\begin{minipage}{0.48\textwidth}
	\centering
	\includegraphics[width=\columnwidth]{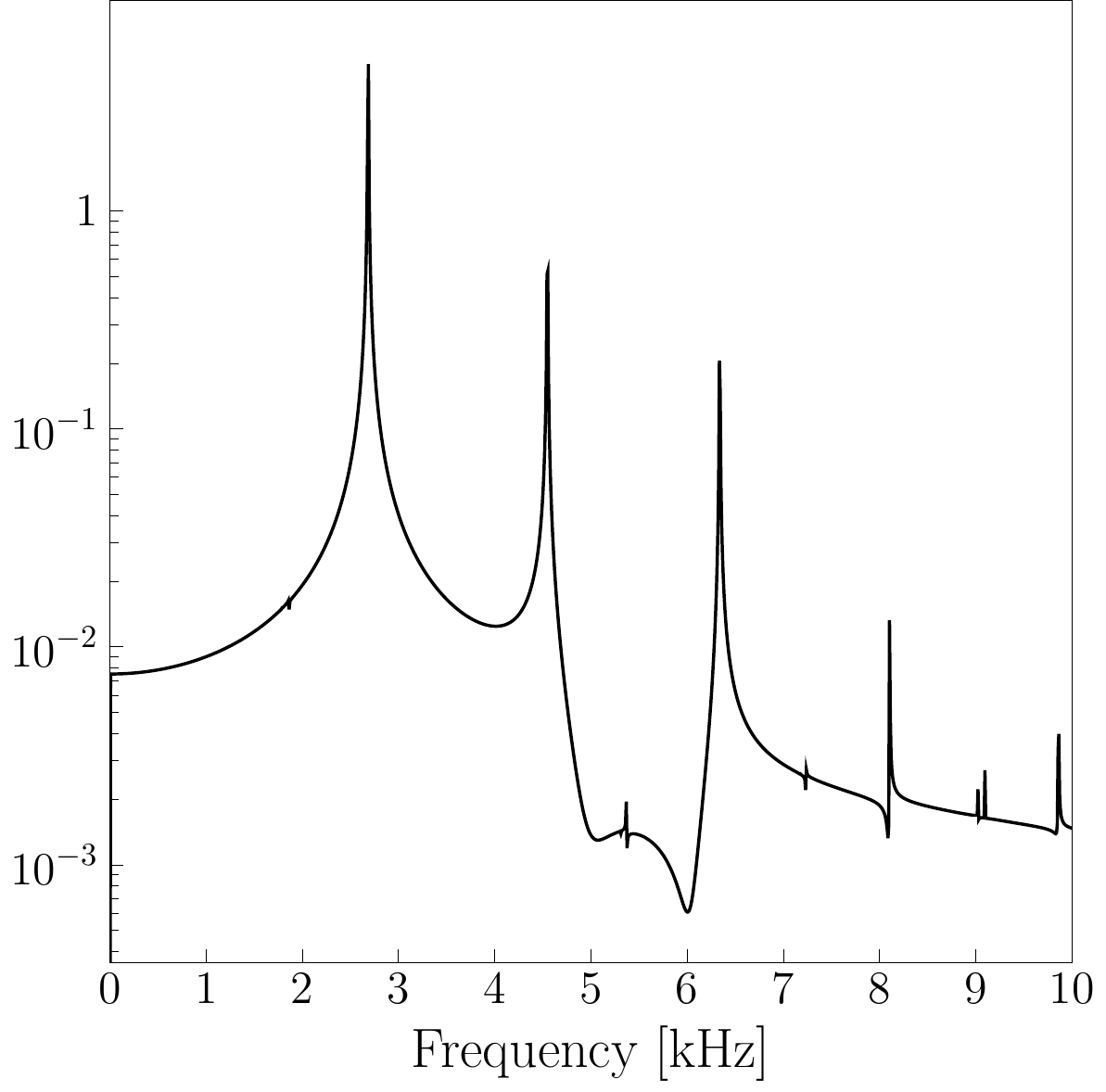}
	\end{minipage}
	\caption{Fourier spectrum of the radius over a 1 s simulation of a $\gamma=2$ polytropic NS with full spacetime evolution (left) and Cowling approximation (right). The frequencies of the three first modes are reported in Tables~\ref{tab:freqpoly} (full spacetime evolution) and~\ref{tab:freqpolycowling} (Cowling approximation). The spectrum is given in arbitrary units.}
	\label{fig:spectrumpoly}
\end{figure}

\begin{figure}
	\centering
	\includegraphics[width=.7\columnwidth]{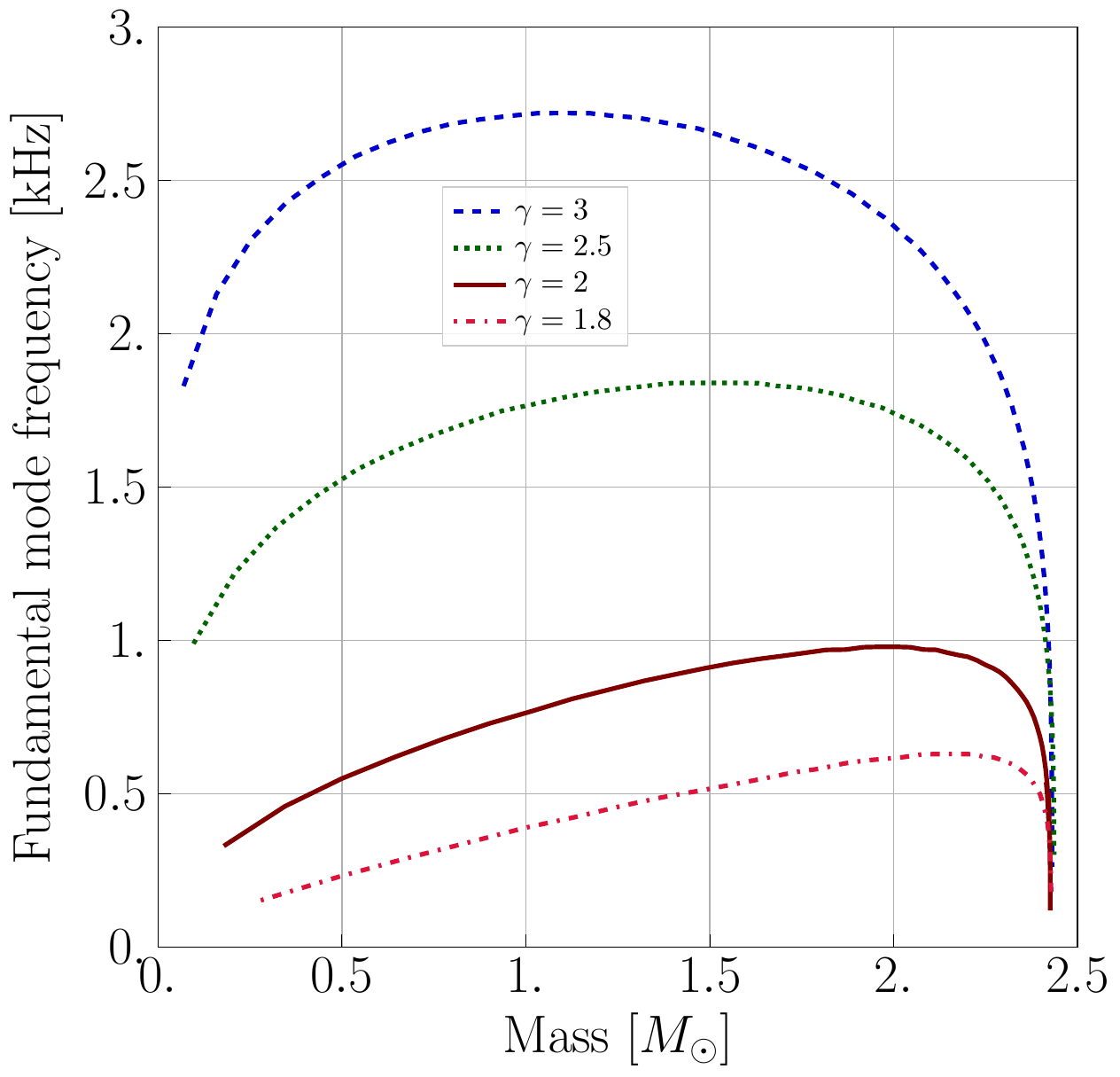}
	\caption{Mass-frequency diagrams for four different polytropes. The $\kappa$ coefficients are $6.23\times10^6$ ($\gamma=1.8$), $4.79\times10^8$ ($\gamma=2$), $1.97\times10^{13}$ ($\gamma=2.5$) and $6.78\times10^{17}$ ($\gamma=3$), in units of $\mathrm{m}^{2(\gamma-1)}$ (in geometrized units i.e. with $G=c=1$).}
	\label{fig:mfdiagpoly}
\end{figure}

\begin{figure}
\centering
\includegraphics[width=.7\columnwidth]{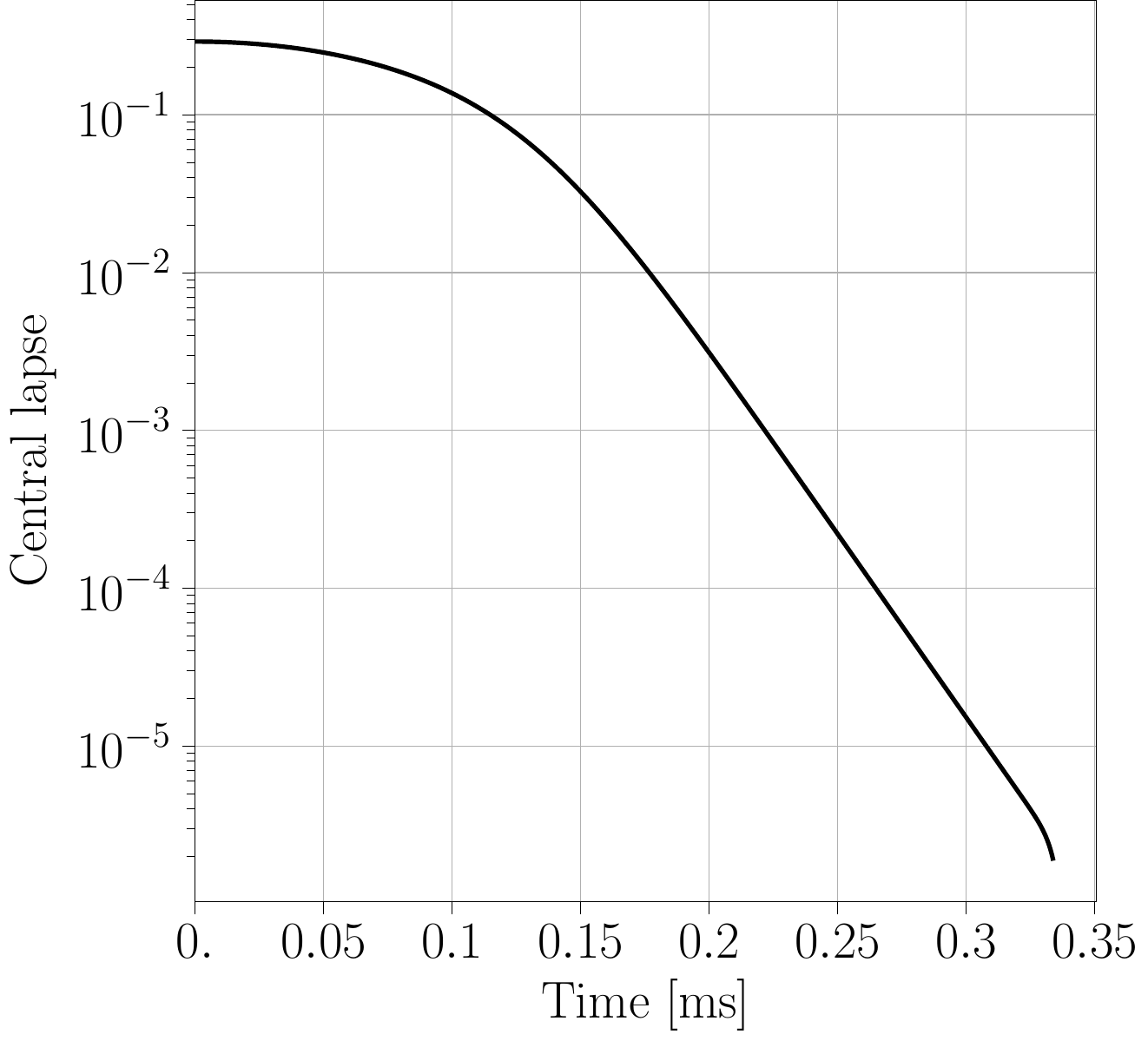}
\caption{BH collapse: lapse function at the center of the star. The collapse of the lapse is characteristic of the formation of a black hole.}
\label{fig:BHcollapselapse}
\end{figure}

\begin{figure}
\centering
\includegraphics[width=.7\columnwidth]{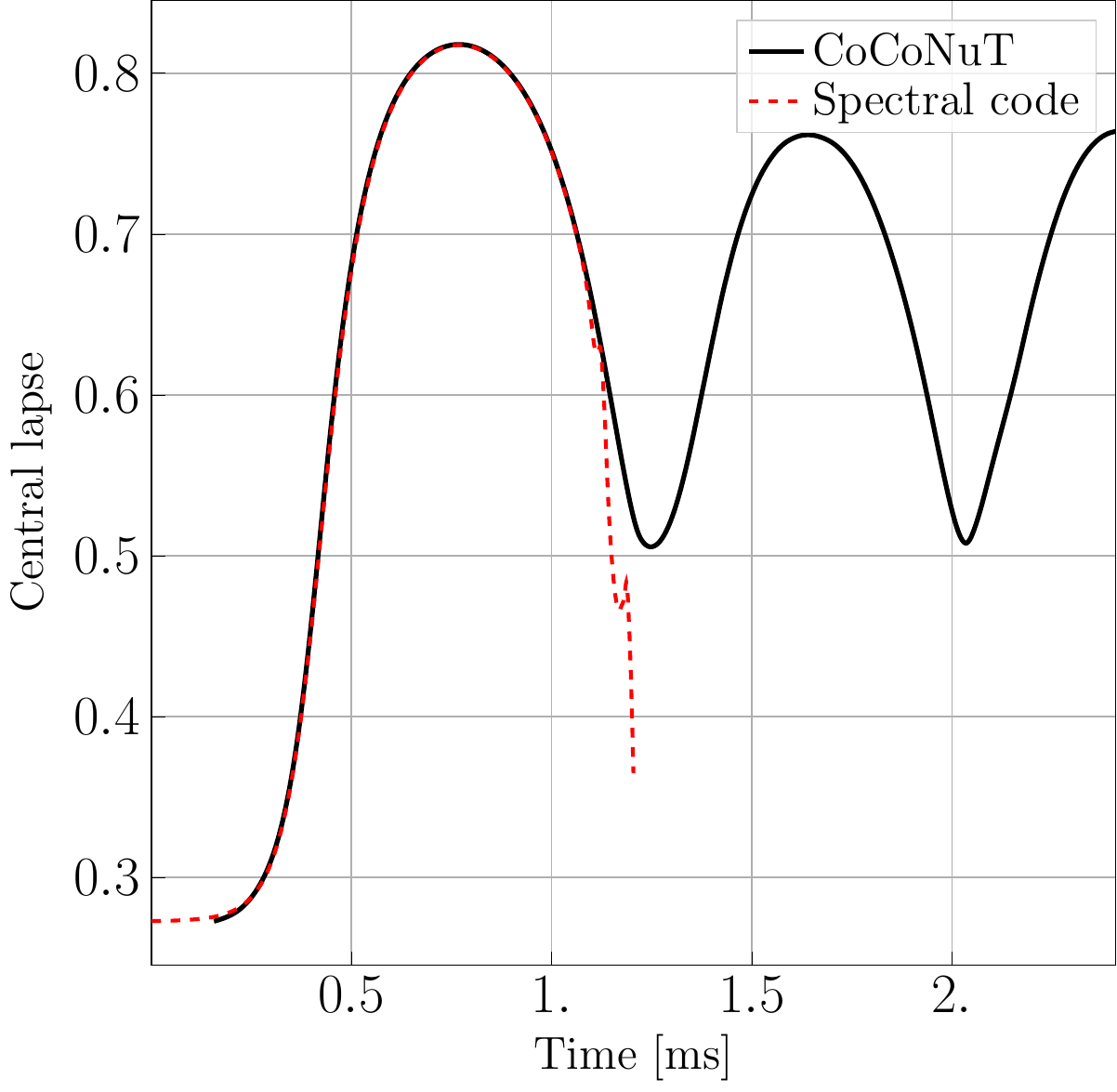}
\caption{Comparing our code with CoCoNuT for a migration test. Please see main text for details.}
\label{fig:migrationtestshock}
\end{figure} 

\subsection{Frequency extraction}\label{subs:frequencyextraction}
We seek to recover the frequencies presented elsewhere in the literature. We use the same star than in Sec.~\ref{subs:convergencetest} and the $\gamma = 5/3,\,\kappa=7.308\,\mathrm{km}^{4/3}$ EoS described in~\cite{hartle_slowly_1975}. The method to extract the frequencies is the following:
\begin{enumerate}
	\item A simulation is run up to some $t_\mathrm{max}$, set to $1 s$ in this test.
	\item A Fast Fourier Transform algorithm is performed on the $R(t)$ signal to compute its spectrum $\hat{R}(f)$.
	\item A local maximum research is performed on $\hat{R}(f)$. When a peak (i.e. a maximum) is detected, the point where the maximum is reached, along with its two neighbours, are interpolated with a quadratic Lagrange polynomial.
	\item The maximum of the polynomial can be analytically extracted; it is an approximation of the frequency of the peak.
\end{enumerate}
The precision on the frequency extraction is less than $1/t_\mathrm{max}$. Note that we compute coordinate frequencies, as the time considered for oscillations is the coordinate time, not the proper time. For the frequencies computed here, we used $N_r=17$ and $\Delta{t} = 3.34\times10^{-4}$ ms. The results are reported in Table~\ref{tab:freqpoly}. The code used in~\cite{font_three-dimensional_2002} is a full GR 3D code with Eulerian high resolution shock capturing methods. The frequencies were also recovered by~\cite{hebert_general-relativistic_2018} with discontinuous Galerkin methods and~\cite{rosswog_sphincs_bssn_2021} with Lagrangian Smooth Particles Hydrodynamics methods. We recover all the frequencies with a precision of less than 1\%. We also look at the frequencies of the~\cite{font_three-dimensional_2002} polytrope in the {Cowling} approximation. The results are reported in Table~\ref{tab:freqpolycowling}. Here we once again recover the tabulated frequencies with an accuracy of less than 1\%. Fig.~\ref{fig:spectrumpoly} shows the spectra for our simulation both in full metric evolution and using the Cowling approximation. On Fig.~\ref{fig:mfdiagpoly} are plotted mass-frequency diagrams for four different polytropes of which the $\kappa$ parameter has been adjusted to yield the same maximum mass, of $2.43\Msol$. To produce such a diagram, we extract frequencies on several stars described by the same EoS, from close to zero mass to maximum mass, by varying the central enthalpy. We see that, as expected~\cite{harrison_gravitation_1972}, the frequency of the fundamental mode of the stars drops down to zero close to the maximum mass. 

We use the occasion to compare our code with the reference core-collapse and NS oscillation code CoCoNuT~\cite{dimmelmeier_combining_2005} which uses pseudospectral methods for the metric equations but which is based upon finite volume methods and a conserved scheme for the hydrodynamics. For the same $\gamma=2$, $\kappa=100$ EoS we used to produce frequencies, a run of the same star, with comparable to worse precision (as measured on the conservation of the ADM mass), is typically five times longer with CoCoNuT than with our code.

\subsection{Collapse to a black hole}\label{subs:blackholecollapse}
The next test we perform is the collapse of a NS to a black hole. The test is rather demanding but often performed~\cite{font_three-dimensional_2002,cordero-carrion_improved_2009,rosswog_sphincs_bssn_2021,thierfelder_numerical_2011,bernuzzi_constraint_2010,baiotti_three_2005}. We use the $\gamma=2$, $\kappa=100$ EoS that we used in Secs.~\ref{subs:convergencetest} and~\ref{subs:frequencyextraction}, and set $H_c=0.61$, on the unstable branch, and perturb it with an enthalpy profile so that it exceeds the maximum mass. The star then collapses into a black hole. We use $N_r=129$ and $\Delta{t} = 3.34\times10^{-6}$ ms. We see on Fig.~\ref{fig:BHcollapselapse} that we indeed find a black hole: the value of the central lapse goes as low as a few $10^{-6}$. We find that the hydrodynamical quantities such as the central pressure, baryon density, log-enthalpy, sound speed squared and the circumferential radius seem to freeze; this is due to the maximal slicing condition, which avoids black hole singularities singularities~\cite{gourgoulhon_31_2012}. The last evidence is the existence of an apparent horizon, which we found using the Apparent Horizon Finder described in~\cite{lin_new_2007}. It appears around $t=0.18\,\mathrm{ms}$ at the corresponding {Schwarzschild} radius of the star: $M_\mathrm{ADM} = 2\Msol$, $R_\mathrm{AH} = 6$ km. This demonstrates the ability of the code to handle very strong gravitational fields.

\subsection{Migration test}

The migration test is the final standard test for numerical isolated NS evolution codes that we perform. We try to reproduce the test originally presented by~\cite{font_three-dimensional_2002} which has also been done by~\cite{cordero-carrion_improved_2009,rosswog_sphincs_bssn_2021,thierfelder_numerical_2011,bernuzzi_constraint_2010,baiotti_three_2005}. Unfortunately, as can be seen in the simulation run with CoCoNuT, a shock is formed and we are able to compute the migration only up to the shock formation, as one would expect from using pseudospectral methods. This is illustrated in Fig.\ref{fig:migrationtestshock} where we compare our simulation with a migration test performed with CoCoNuT~\cite{dimmelmeier_combining_2005}. The simulation is run with $N_r=17$ and $\Delta{t}=3.34\times10^{-4}$. The migration being triggered by numerical truncation error, it has no reason to start at the same time, so we have to shift the time of the one performed with CoCoNuT to align the curves. We see that because our simulation develops Gibbs phenomenon, the star is not able to oscillate. It is worth noting that, when run on a laptop, our codes reaches the point where it crashes in approximately ten seconds, while CoCoNuT takes six minutes to reach that same point.

\section{Summary and conclusions}\label{s:conclusions}
We have presented a new formulation of the hydrodynamical equations in General Relativity, using the 3+1 formalism, aimed for numerical simulations. Those equations only rely on the primitive variables, meaning that, as opposed to the widely used conservative form of the Valencia hydrodynamics formalism~\cite{banyuls_numerical_1997}, no recovery procedure is needed to compute the sources for the metric equations. The derivation of the equation relies on a procedure that shows the evolution of the variables $n_X,\,S,\,U_i,\,\Gamma,\,\gamma$ and $p$ is actually degenerate and decouples them to end up with equations for $n_X,\,S$ and $U_i$ only. This formulation allows for the simulation of any relativistic perfect fluid, not only self-gravitating objects, and is expressed in a 3+1 covariant form. Then, we have summarized the equations for a fluid with an EoS that uses at most three parameters. The extension to more parameters, for example accounting for the presence of muons, is straightforward as the new equations needed are already encoded in the general derivation. We then applied them to the simulation of NS oscillations.
The code used for numerical tests implements a fully constrained formalism for the computation of the metric. It uses an Adams-Bashforth finite-differences scheme for the time evolution, and pseudospectral methods for the spatial coordinates. Pseudospectral methods are well-suited for smooth flows and help to save even more computational resources. A notable feature is that, like in the Lagrangian approach of~\cite{rosswog_sphincs_bssn_2021}, the surface of the star does not need any specific treatment; we use the fact that it is well-defined at equilibrium, and then evolve the grid according to the star evolution. The code is still at an early stage of development, hence only spherically symmetric configurations have been implemented. Nonetheless, the frequencies of polytropic NS are recovered with excellent precision, and the code is also able to perform a black hole collapse. Only the migration test was not fully run because of a shock formation, which also highlights the limitations of our numerical approach. Future work includes the extension of the code to three spatial dimensions, as well as two or three parameters in the EoS.

\ack
The authors gratefully acknowledge the Italian Instituto Nazionale de Fisica Nucleare (INFN), the French Centre National de la Recherche Scientifique (CNRS) and the Netherlands Organization for Scientific Research for the construction and operation of the Virgo detector and the creation and support of the EGO consortium. 
This work has been supported by the Spanish Agencia Estatal de Investigaci\'on / 
Ministerio de Ciencia, Innovaci\'on y Universidades through the Grants No. 
PGC2018-095984-B-I00 and PID2021-125458NB-C21, the Generalitat Valenciana through the Grant No. 
PROMETEO/2019/071, the “Action pluriannuelle incitative ondes gravitationnelles et objets compacts de l'Observatoire de Paris” and the CNRS International Research Project (IRP) “Origine des éléments lourds dans l’univers: Astres Compacts et Nucléosynthèse (ACNu)”. The authors warmly thank \'Eric Gourgoulhon and Micaela Oertel for having carefully read the manuscript and helped improve it.

\appendix
\section{Characteristic speed for the conformal factor evolution} \label{app:characteristic}
Eq.~(\ref{eq:evolpsi}) is an advection equation, therefore taking the moving grid into account, the characteristic speed is
    \begin{equation}
        -(\beta^r(R(t),t)+\dot{R}(t)\xi)
    \end{equation}
    in the nucleus and 
    \begin{equation}
        (\xi-1)\dot{R}(t) - \frac{(\xi-1)^2}{2}\beta^r(R(t),t)
    \end{equation}
    in the CED. At the exterior boundary of the nucleus, the characteristic speed reads
    \begin{equation}
        -\beta^r(R(t),t) - \dot{R}(t) = -NU^r(R(t),t),
    \end{equation}
    and at the interior boundary of the CED, the characteristic speed is
    \begin{equation}
        -2(\dot{R}(t) + \beta^r(R(t),t)) = -2NU^r(R(t),t).
    \end{equation}
    Therefore, the lapse function being positive, when $U^r(R(t),t)<0$, the value of $\ln\Psi$ at the last grid point of the nucleus is copied to the first grid point of the CED, and \textit{vice-versa} for $U^r(R(t),t)>0$.

\section*{References}
\bibliographystyle{unsrt}
\bibliography{biblio}

\end{document}